# Revisiting Centrality-as-Relevance: Support Sets and Similarity as Geometric Proximity

**Ricardo Ribeiro**  RICARDO.RIBEIRO@INESC-ID.PT
*Instituto Universitário de Lisboa (ISCTE-IUL)*
*Av. das Forças Armadas, 1649-026 Lisboa, Portugal*
*L2F - INESC ID Lisboa*
*Rua Alves Redol, 9, 1000-029 Lisboa, Portugal*

**David Martins de Matos**  DAVID.MATOS@INESC-ID.PT
*Instituto Superior Técnico, Universidade Técnica de Lisboa*
*Av. Rovisco Pais, 1049-001 Lisboa, Portugal*
*L2F - INESC ID Lisboa*
*Rua Alves Redol, 9, 1000-029 Lisboa, Portugal*

## Abstract

In automatic summarization, centrality-as-relevance means that the most important content of an information source, or a collection of information sources, corresponds to the most central passages, considering a representation where such notion makes sense (graph, spatial, etc.). We assess the main paradigms, and introduce a new centrality-based relevance model for automatic summarization that relies on the use of support sets to better estimate the relevant content. Geometric proximity is used to compute semantic relatedness. Centrality (relevance) is determined by considering the whole input source (and not only local information), and by taking into account the existence of minor topics or lateral subjects in the information sources to be summarized. The method consists in creating, for each passage of the input source, a support set consisting only of the most semantically related passages. Then, the determination of the most relevant content is achieved by selecting the passages that occur in the largest number of support sets. This model produces extractive summaries that are generic, and language- and domain-independent. Thorough automatic evaluation shows that the method achieves state-of-the-art performance, both in written text, and automatically transcribed speech summarization, including when compared to considerably more complex approaches.

## 1. Introduction

A summary conveys to the end user the most relevant content of one or more information sources, in a concise and comprehensible manner. Several difficulties arise when addressing this problem, but one of utmost importance is how to assess the significant content. Usually, approaches vary in complexity if processing text or speech. While in text summarization, up-to-date systems make use of complex information, such as syntactic (Vanderwende, Suzuki, Brockett, & Nenkova, 2007), semantic (Tucker & Spärck Jones, 2005), and discourse information (Harabagiu & Lacatusu, 2005; Uzêda, Pardo, & Nunes, 2010), either to assess relevance or reduce the length of the output, common approaches to speech summarization try to cope with speech-related issues by using speech-specific information (for example, prosodic features, Maskey & Hirschberg, 2005, or recognition confidence scores, Zechner & Waibel, 2000) or by improving the intelligibility of the output of an automatic speech recognition system (by using related information, Ribeiro & de Matos, 2008a). In fact, spoken language summarization is often considered a much harder task than text summarization (McKeown, Hirschberg, Galley, & Maskey, 2005; Furui, 2007): problems like speech recognition errors, disfluencies, and the accurate identification of sentence boundaries not only increase the difficulty in determining the salient information, but also constrain the applicability of text summarization techniques to speech summarization (although in the presence of planned speech, as it partly happens in the broadcast news domain, that portability is more feasible, Christensen, Gotoh, Kolluru, & Renals, 2003). Nonetheless, shallow text summarization approaches such as Latent Semantic Analysis (LSA) (Landauer, Foltz, & Laham, 1998; Gong & Liu, 2001) and Maximal Marginal Relevance (MMR) (Carbonell & Goldstein, 1998) seem to achieve performances comparable to the ones using specific speech-related features (Penn & Zhu, 2008).

Following the determination of the relevant content, the summary must be composed and presented to the user. If the identified content consists of passages found in the input source that are glued together to form the summary, that summary is usually designated as *extract*; on the other hand, when the important content is devised as a series of concepts that are fused into a smaller set and then used to generate a new, concise, and informative text, we are in the presence of an *abstract*. In between extraction and concept-to-text generation, especially in text summarization, text-to-text generation methods, which rely on text rewriting—paraphrasing—, of which sentence compression is a major representative, are becoming an up-to-date subject (Cohn & Lapata, 2009). Given the hardness of abstraction, the bulk of the work in the area consists of extractive summarization.

A common family of approaches to the identification of the relevant content is the *centrality* family. These methods base the detection of the most salient passages on the identification of the central passages of the input source(s). One of the main representatives of this family is *centroid-based summarization*. Centroid-based methods build on the idea of a pseudo-passage that represents the central topic of the input source (the *centroid*) selecting as passages ($x$) to be included in the summary the ones that are close to the centroid. Pioneer work (on multi-document summarization) by Radev, Hatzivassiloglou, and McKeown (1999) and Radev, Jing, and Budzikowska (2000) creates clusters of documents by representing each document as a *tf-idf* vector; the centroid of each cluster is also defined as a *tf-idf* vector, with the coordinates corresponding to the weighted average of the *tf-idf* values of the documents of the cluster; finally, sentences that contain the words of the centroids are presumably the best representatives of the topic of the cluster, thus being the best candidates to belonging to the summary.

$$centrality(x) = similarity(x, centroid) \quad (1)$$

Another approach to centrality estimation is to compare each candidate passage to every other passage ($y$) and select the ones with higher scores (the ones that are closer to every other passage). One simple way to do this is to represent passages as vectors using a weighting scheme like the aforementioned *tf-idf*; then, passage similarity can be assessed using, for instance, the cosine, assigning to each passage a centrality score as defined in Eq. 2.

$$centrality(x) = \frac{1}{N} \sum_y similarity(x, y) \quad (2)$$

These scores are then used to create a sentence ranking: sentences with highest scores are selected to create the summary.

A major problem of this relevance paradigm is that by taking into account the entire input source in this manner, either to estimate centroids or average distances of input source passages, we may be selecting extracts that being *central* to the input source are, however, not the most relevant ones. In cognitive terms, the information reduction techniques in the summarization process are quite close to the discourse understanding process (Endres-Niggemeyer, 1998), which, at a certain level, works by applying rules that help uncovering the macrostructure of the discourse. One of these rules, *deletion*, is used to eliminate from the understanding process propositions that are not relevant to the interpretation of the subsequent ones. This means that it is common to find, in the input sources to be summarized, lateral issues or considerations that are not relevant to devise the salient information (discourse structure-based summarization is based on the relevance of nuclear text segments, Marcu, 2000; Uzêda et al., 2010), and that may affect centrality-based summarization methods by inducing inadequate centroids or decreasing the scores of more suitable sentences.

As argued by previous work (Gong & Liu, 2001; Steyvers & Griffiths, 2007), we also assume that input sources are mixtures of topics, and propose to address that aspect using the input source itself as guidance. By associating to each passage of the input source a support set consisting only of the most semantically related passages in the same input source, groups of related passages are uncovered, each one constituting a latent topic (the union of the supports sets whose intersection is not empty). In the creation of these support sets, semantic relatedness is assessed by geometric proximity. Moreover, while similar work usually explores different weighting schemes to address specific issues of the task under research (Orăsan, Pekar, & Hasler, 2004; Murray & Renals, 2007; Ribeiro & de Matos, 2008b), we explore different geometric distances as similarity measures, analyzing their performance in context (the impact of different metrics from both theoretical and empirical perspectives in a clustering setting was shown in Aggarwal, Hinneburg, & Keim, 2001). To build the summary, we select the sentences that occur in the largest number of support sets—hence, the most *central* sentences, without the problem that affects previous centrality-based summarization.

Our method produces generic, language- and domain-independent summaries, with low computational requirements. We test our approach both in speech and text data. In the empirical evaluation of the model over text data, we used an experimental setup previously used in published work (Mihalcea & Tarau, 2005; Antiqueira, Oliveira Jr., da Fontoura Costa, & Nunes, 2009), which enabled an informative comparison to the existing approaches. In what concerns the speech experiments, we also used a corpus collected in previous work (Ribeiro & de Matos, 2008a), as well as the published results. This allowed us to compare our model to state-of-the-art work.

The rest of this document is structured as follows: in Section 2, we analyze representative models of both centrality-as-relevance approaches—passage-to-centroid similarity-based centrality and pair-wise passage similarity-based centrality; Section 3 describes the support sets-based relevance model; the evaluation of the model is presented in Section 4, where we compare its performance against other centrality-as-relevance models and discuss the achieved results; final remarks conclude the document.

## 2. Centrality-as-Relevance

There are two main approaches to centrality-based summarization: passage-to-centroid similarity and pair-wise passage similarity.

### 2.1 Passage-to-Centroid Similarity-based Centrality

In centroid-based summarization, passage centrality is defined by the similarity between the passage and a pseudo-passage that, considering a geometrical representation of the input source, is the center of the space defined by the passages of the input source, the centroid. The work in multi-document summarization by Radev et al. (1999, 2000) and Radev, Jing, Styś, and Tam (2004) and the work developed by Lin and Hovy (2000) are examples of this approach.

Radev et al. present a centroid-based multi-document summarizer (MEAD) that has as input a cluster of documents. Associated to each cluster of documents is a centroid. Documents are represented by vectors of *tf-idf* weights and the centroid of each cluster consists of a vector which coordinates are the weighted averages of the *tf-idf* values of the documents of the cluster, above a pre-defined threshold. Thus, the centroid of a cluster of documents is, in this case, a pseudo-document composed by the terms that are statistically relevant. Given a cluster of documents segmented into sentences $IS \triangleq \{s_1, s_2, \ldots, s_N\}$, a centroid $C$, and a compression rate, summarization is done by selecting the appropriate number (according to the compression rate) of the sentences with the highest scores assigned by a linear function (Eq. 3) of the following features: *centroid value* ($C_i$), *position value* ($P_i$), and *first-sentence overlap value* ($F_i$).

$$score(s_i) = w_c C_i + w_p P_i + w_f F_i, \quad 1 \leq i \leq N \tag{3}$$

The centroid value, defined as $C_i = \sum_{t \in s_i} C_{t,i}$, establishes that the sentences closer to the centroid (the ones that contain more terms $t$ from the centroid) have higher scores. Position value ($P_i$) scores sentences according to their position in the encompassing document. Finally, first-sentence overlap value ($F_i$) scores sentences according to their similarity to the first sentence of the document.

Lin and Hovy (2000) designate the centroid as *topic signature* and define it as a set of related terms: $TS \triangleq \{topic, <(t_1, w_1), \ldots, (t_T, w_T)>\}$, where $t_i$ represents a term related to the topic *topic* and $w_i$ is an associated weight that represents the degree of correlation of $t_i$ to the topic. Topic signatures are computed from a corpus of documents, previously classified as relevant or non-relevant for a given topic, using the log-likelihood-ratio-based quantity $-2log(\lambda)$. This quantity, due to its asymptotic relation to the $\chi^2$ distribution as well as the adequacy of the log-likelihood-ratio to sparse data, is used to rank the terms that will define the signature, and to select a cut-off value that will establish the number of terms in the signature. Summarization is carried out by ranking the sentences according to the topic signature score and selecting the top ranked ones. The topic signature score ($tss$) is computed in a similar manner to MEAD's centroid value: given an input source $IS \triangleq \{p_1, p_2, \ldots, p_N\}$, where $p_i \triangleq \langle t_1, \ldots, t_M \rangle$, the most relevant passages are the ones with more words from the topic (Eq. 4).

$$tss(p_i) = \sum_{j=1}^{M} w_j, \text{with } w_j \text{ weight of } t_j \text{ as defined in a topic signature} \tag{4}$$

## 2.2 Pair-wise Passage Similarity-based Centrality

In pair-wise passage similarity-based summarization, passage centrality is defined by the similarity between each passage and every other passage. The work presented by Erkan and Radev (2004), as well as the work developed by Mihalcea and Tarau (2005), are examples of this approach.

Erkan and Radev (2004) propose three graph-based approaches to pair-wise passage similarity-based summarization with similar performance: degree centrality, LexRank, and continuous LexRank. Degree centrality is based on the degree of a vertex. Pair-wise sentence similarity is used to build a graph representation of the input source: vertices are sentences and edges connect vertices which corresponding sentences are similar above a given threshold. Sentences similar to a large number of other sentences are considered the most central (relevant) ones. Degree centrality is similar to the model we propose. However, in the model we propose, we introduce the concept of support set to allow the use of a different threshold for each sentence. This improves the representation of each sentence, leading to the creation of better summaries.

LexRank, based on Google's PageRank (Brin & Page, 1998), builds on degree centrality (*degree*) by making the centrality of a sentence $s$ be influenced by similar sentences, the adjacent ones in the graph representation (Eq. 5).

$$centralityScore(s) = \sum_{t \in adj[s]} \frac{centralityScore(t)}{degree(t)} \tag{5}$$

The ranking model is similar to PageRank except in what concerns the similarity (adjacency) graph, that, in this case, is undirected (Eq. 6, $d$ is a damping factor and $N$ the number of sentences).

$$centrality(s) = \frac{d}{N} + (1-d) \sum_{t \in adj[s]} \frac{centrality(t)}{degree(t)} \tag{6}$$

Continuous LexRank is a weighted version of LexRank (it uses Eq. 7 instead of Eq. 5).

$$centralityScore(s) = \sum_{t \in adj[s]} \frac{sim(s,t)}{\sum_{u \in adj[t]} sim(u,t)} centralityScore(t) \tag{7}$$

Mihalcea and Tarau (2005), in addition to Google's PageRank, also explore the HITS algorithm (Kleinberg, 1999) to perform graph-based extractive text summarization: again, documents are represented as networks of sentences and these networks are used to globally determine the importance of each sentence. As it happens in the models proposed by Erkan and Radev, sentences are vertices ($V$) and edges ($w$) between vertices are established by passage similarity. The TextRank (Mihalcea & Tarau, 2004)—how the model based on PageRank was designated and the main contribution—formalization is similar to Continuous LexRank (see Eq. 8), although Mihalcea and Tarau also explore directed graphs in the representation of the text[12]. For summarization, the best results were obtained using a backward directed graph: the orientation of the edges from a vertex representing a sentence is to vertices representing previous sentences in the input source.

$$TextRank(V_s) = (1-d) + d * \sum_{V_t \in In[V_s]} \frac{w(V_t, V_s)}{\sum_{V_u \in Out[V_t]} w(V_t, V_u)} TextRank(V_t) \qquad (8)$$

Passage similarity is based on content overlap[3] and is defined in Eq. 9. Given two sets $P \triangleq p_1, p_2, ..., p_n$ and $Q \triangleq q_1, q_2, ..., q_n$, each corresponding to a passage, similarity consists in the cardinality of the intersection over the sum of the logarithms of the cardinality of each set.

$$w(V_P, V_Q) = sim(P, Q) = \frac{|\{t : t \in P \wedge t \in Q\}|}{\log(|P|) + \log(|Q|)} \qquad (9)$$

A similar graph-based approach is described by Antiqueira et al. (2009). This work uses complex networks to perform extractive text summarization. Documents are also represented as networks, where the sentences are the nodes and the connections between nodes are established between sentences sharing common meaningful nouns.

### 2.3 Beyond Automatic Summarization

Apart from summarization, and considering that PageRank and HITS stem from the area of Information Retrieval, centrality-based methods similar to the ones previously described have been successfully applied to re-rank sets of documents returned by retrieval methods.

Kurland and Lee (2005, 2010) present a set of graph-based algorithms, named *influx*, that are similar to our model, to reorder a previously retrieved collection of documents ($C$). The method starts by defining a $k$-nearest-neighbor ($k$NN) graph over the initial collection based on generation links defined as in Eq. 10 ($KL$, Kullback-Leibler divergence; $MLE$, maximum-likelihood estimate; $\mu$, smoothing-parameter of a Dirichlet-smoothed version of $\tilde{p}(\cdot)$; $d$ and $s$, documents).

$$p_d^{KL,\mu}(s) \triangleq \exp\left(-KL\left(\tilde{p}_s^{MLE}(\cdot) \,\middle\|\, \tilde{p}_d^{[\mu]}(\cdot)\right)\right) \qquad (10)$$

Centrality is determined as defined in Eq. 11. Edges can be weighted (weight given by $p_d^{KL,\mu}(s)$) or not (weight is 1). Edges corresponding to generation probabilities below the $k$ highest ones are not considered.

$$centralityScore(d) \triangleq \sum_{o \in C} wt(o \to d) \qquad (11)$$

---

1. In "A Language Independent Algorithm for Single and Multiple Document Summarization" (Mihalcea & Tarau, 2005), the weighted PageRank equation has a minor difference from the one in "TextRank: Bringing Order into Texts" (Mihalcea & Tarau, 2004). The latter presents the correct equation.
2. Although both LexRank and TextRank are based on PageRank, different equations are used in their formalization. The equation used in TextRank formalization is the same of PageRank original publication, however PageRank authors observe that *the PageRanks form a probability distribution over Web pages, so the sum of all Web pages' PageRanks will be one*. This indicates the need of the normalization factor that is observed in LexRank formalization and currently assumed to be the correct PageRank formalization.
3. The metric proposed by Mihalcea and Tarau (2004) has an unresolved issue: the denominator is 0 when comparing two equal sentences with length one (something that can happen when processing speech transcriptions). Instead, the Jaccard similarity coefficient (1901) could be used.

There are also recursive versions of this centrality model, which are similar to PageRank/LexRank and Continuous LexRank.

## 3. Support Sets and Geometric Proximity

In this work, we hypothesize that input sources to be summarized comprehend different topics (lateral issues beyond the main topic), and model this idea by defining a support set, based on semantic relatedness, for every passage in the input source. Semantic relatedness is estimated within the geometric framework, where we explore several distance metrics to compute proximity. The most relevant content is determined by computing the most central passages given the collection of support sets. The proposed model estimates the most salient passages of an input source, based exclusively on information drawn from the used input source.

### 3.1 Model

The leading concept in our model is the concept of support set: the first step of our method to assess the relevant content is to create a support set for each passage of the input source by computing the similarity between each passage and the remaining ones, selecting the closest passages to belong to the support set. The most relevant passages are the ones that occur in the largest number of support sets.

Given a segmented information source $I \triangleq p_1, p_2, ..., p_N$, support sets $S_i$ associated with each passage $p_i$ are defined as indicated in Eq. 12 ($sim()$ is a similarity function, and $\varepsilon_i$ is a threshold).

$$S_i \triangleq \{s \in I : sim(s, p_i) > \varepsilon_i \wedge s \neq p_i\} \tag{12}$$

The most relevant segments are given by selecting the passages that satisfy Eq. 13.

$$\underset{s \in \cup_{i=1}^n S_i}{\arg\max} \left| \{S_i : s \in S_i\} \right| \tag{13}$$

A major difference from previous centrality models and the main reason to introduce the support sets is that by allowing different thresholds to each set ($\varepsilon_i$), we let centrality be influenced by the latent topics that emerge from the groups of related passages. In the degenerate case where all $\varepsilon_i$ are equal, we fall into the degree centrality model proposed by Erkan and Radev (2004). But using, for instance, a naïve approach of having dynamic thresholds ($\varepsilon_i$) set by limiting the cardinality of the support sets (a $k$NN approach), centrality is changed because each support set has only the most semantically related passages of each passage. From a graph theory perspective, this means that the underlying representation is not undirected, and the support set can be interpreted as the passages recommended by the passage associated to the support set. This contrasts with both LexRank models, which are based on undirected graphs. On the other hand, the models proposed by Mihalcea and Tarau (2005) are closer to our work in the sense that they explore directed graphs, although only in a simple way (graphs can only be directed forward or backward). Nonetheless, semantic relatedness (content overlap) and centrality assessment (performed by the graph ranking algorithms HITS and PageRank) is quite different from our proposal. In what concerns the work of Kurland and Lee (2005, 2010), which considering this $k$NN approach to the definition of the support set size, is the most similar to our ideias, although not addressing automatic summarization, the neighborhood definition strategy is different than ours: Kurland and Lee base neighborhood definition on generation probabilities (Eq. 10), while we explore geometric proximity. Nevertheless, from the perspective of our model, the $k$NN approach to support set definition is only a possible strategy (others can be used): our model can be seen as a generalization of both $k$NN and $\varepsilon$NN approaches, since what we propose is the use of differentiated thresholds ($\varepsilon_i$) for each support set (Eq. 12).

### 3.2 Semantic Space

We represent the input source $I$ in a term by passages matrix $A$, where each matrix element $a_{ij} = f(t_i, p_j)$ is a function that relates the occurrences of each term $t_i$ within each passage $p_j$ ($T$ is the number of different terms; $N$ is the number of passages).

$$A = \begin{bmatrix} a_{1,1} & \ldots & a_{1,N} \\ \ldots & & \\ a_{T,1} & \ldots & a_{T,N} \end{bmatrix} \tag{14}$$

In what concerns the definition of the weighting function $f(t_i, p_j)$, several term weighting schemes have been explored in the literature—for the analysis of the impact of different weighting schemes on either text or speech summarization see the work of Orăsan et al. (2004), and Murray and Renals (2007) or Ribeiro and de Matos (2008b), respectively. Since the exact nature of the weighting function, although relevant, is not central to our work, we opted for normalized frequency for simplicity, as defined in Eq. 15, where $n_{i,j}$ is the number of occurrences of term $t_i$ in passage $p_j$.

$$f(t_i, p_j) = tf_i = \frac{n_{i,j}}{\sum_k n_{k,j}} \tag{15}$$

Nevertheless, this is in line with the work of Sahlgren (2006) that shows that in several tasks concerning term semantic relatedness, one of the most effective weighting schemes for small contexts is the binary term weighting scheme (Eq. 16), alongside raw or dampened counts, that is, weighting schemes, based on the frequency, that do not use global weights (note also that in such small contexts, most of the words have frequency 1, which normalized or not is similar to the binary weighting scheme).

$$f(t_i, p_j) = \begin{cases} 1 & \text{if } t_i \in p_j \\ 0 & \text{if } t_i \notin p_j \end{cases} \tag{16}$$

### 3.3 Semantic Relatedness

As indicated by Sahlgren (2006), *the meanings-are-locations metaphor is completely vacuous without the similarity-is-proximity metaphor*. In that sense, we explore the prevalent distance measures found in the literature, based on the general Minkowski distance (Eq. 17).

$$dist_{minkowski}(\mathbf{x}, \mathbf{y}) = \left( \sum_{i=1}^{n} |x_i - y_i|^N \right)^{\frac{1}{N}} \tag{17}$$

Semantic relatedness is computed using the Manhattan distance ($N = 1$, Eq. 19), the Euclidean distance ($N = 2$, Eq. 20), the Chebyshev distance ($N \to \infty$, Eq. 21), and fractional distance metrics (we experimented with $N = 0.1$, $N = 0.5$, $N = 0.75$, and $N = 1.(3)$. Note that, when $0 < N < 1$, Eq. 17 does not represent a metric, since the triangle inequality does not hold (Koosis, 1998, page 70). In this case, it is common to use the variation defined in Eq. 18.

$$dist_N(\mathbf{x}, \mathbf{y}) = \sum_{i=1}^{n} |x_i - y_i|^N,\ 0 < N < 1 \tag{18}$$

Moreover, we also experiment with the general Minkowski equation, using the tuple dimension as $N$.

$$dist_{manhattan}(\mathbf{x}, \mathbf{y}) = \sum_{i=1}^{n} |x_i - y_i| \tag{19}$$

$$dist_{euclidean}(\mathbf{x}, \mathbf{y}) = \sqrt{\sum_{i=1}^{n} (x_i - y_i)^2} \tag{20}$$

$$dist_{chebyshev}(\mathbf{x}, \mathbf{y}) = \lim_{N \to \infty} \Big( \sum_{i=1}^{n} |x_i - y_i|^N \Big)^{\frac{1}{N}} = \max_{i} \left(|x_i - y_i|\right) \tag{21}$$

The cosine similarity (Eq. 22), since it is one of the most used similarity metrics, especially when using spatial metaphors for computing semantic relatedness, was also part of our experiments.

$$sim_{\cos}(\mathbf{x}, \mathbf{y}) = \frac{\mathbf{x} \cdot \mathbf{y}}{\|\mathbf{x}\|\|\mathbf{y}\|} = \frac{\sum_{i=1}^{n} x_i y_i}{\sqrt{\sum_{i=1}^{n} x_i^2} \sqrt{\sum_{i=1}^{n} y_i^2}} \tag{22}$$

Grounding semantic relatedness on geometric proximity enables a solid analysis of the various similarity metrics. For instance, when using the Euclidean distance (Eq. 20), differences between tuple coordinate values less than 1 make passages closer, while values greater than 1 make passages more distant; Chebyshev's distance (Eq. 21) only takes into account one coordinate: the one with the greatest difference between the two passages; and, the Manhattan distance (Eq. 19) considers all coordinates evenly. In the cosine similarity (Eq. 22), tuples representing passages are vectors and the angle they form establishes their relatedness. In contrast, Mihalcea and Tarau (2005) and Antiqueira et al. (2009) define passage similarity as content overlap. Figure 1 ($N$ ranges from 0.1, with an almost imperceptible graphical representation, to $N \to \infty$, a square) shows how the unit circle is affected by the several geometric distances (Manhattan, $N = 1$, and Euclidean, $N = 2$, are highlighted).

Although geometric proximity enables a solid analysis of the effects of using a specific metric, it mainly relies on lexical overlap. Other metrics could be used, although the costs in terms of the required resources would increase. Examples are corpus-based vector space models of semantics (Turney & Pantel, 2010), like LSA (Landauer et al., 1998), Hyperspace Analogue to Language (Lund, Burgess, & Atchley, 1995), or Random Indexing (Kanerva, Kristoferson, & Holst, 2000; Kanerva & Sahlgren, 2001), or similarity metrics based on knowledge-rich semantic resources, such as WordNet (Fellbaum, 1998).

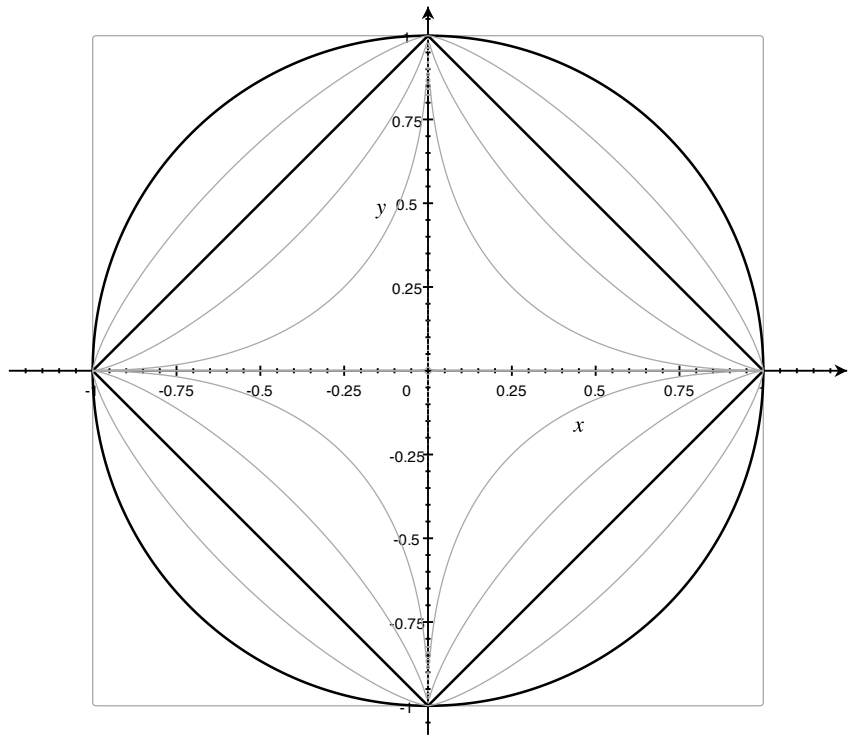

Figure 1: Unit circles using various fractional distance metrics ($N$ equals to 0.1, 0.5, 0.75, and 1.(3)), the Manhattan distance ($N = 1$), the Euclidean distance ($N = 2$), and the Chebyshev's distance ($N \to \infty$).

### 3.4 Threshold Estimation

As previously mentioned, a simple approach to threshold estimation is to define a fixed cardinality for all support sets, a $k$NN approach. This means that thresholds, although unknown, are different for each support set.

A simple heuristic that allows to automatically set per passage thresholds is to select as members of the support set the passages which distance to the passage associated to the support set under construction is smaller than the average distance. In the next sections, we explore several heuristics inspired by the nature of the problem that can be used as possibly better approaches to threshold estimation. However, this subject merits further study.

3.4.1 Heuristics Based on Distance Progression Analysis

One possible approach is to analyze the progression of the distance values between each passage and the remaining ones in the creation of the respective support set. This type of heuristics uses a sorted permutation, $d^i_1 \leq d^i_2 \leq \cdots \leq d^i_{N-1}$, of the distances of the passages, $s_k$, to the passage $p_i$ (corresponding to the support set under construction), with $d^i_k = dist(s_k, p_i)$, $1 \leq k \leq N-1$, and $N$ the number of passages.

We explore three approaches: a standard deviation-based approach, where $\varepsilon_i$ is given by Eq. 23, with $\alpha$ a parameter that controls the width of interval around the average distance in relation to the standard deviation; an approach based on the diminishing differences between consecutive distances, $d^i_{k+2} - d^i_{k+1} < d^i_{k+1} - d^i_k$, $1 \leq k \leq N-3$, where $\varepsilon_i = d^i_{k+2}$, such that $k$ is the largest one that $\forall_{1 \leq j \leq k+1} j : d^i_{j+2} - d^i_{j+1} < d^i_{j+1} - d^i_j$; and, an approach based on the average difference between consecutive distances, $d^i_{k+1} - d^i_k < \frac{\sum_{l=1}^{N-2}(d^i_{l+1} - d^i_l)}{N-2}$, $1 \leq k \leq N-2$, where $\varepsilon_i = d_{i_{k+1}}$, such that $k$ is the largest one that $\forall_{1 \leq j \leq k} j : d^i_{j+1} - d^i_j < \frac{\sum_{l=1}^{N-2}(d^i_{l+1} - d^i_l)}{N-2}$.

$$\varepsilon_i = \mu_i - \alpha \sigma_i, \text{ with} \qquad (23)$$

$$\mu_i = \frac{1}{N-1} \sum_{k=1}^{N-1} d^i_k, \text{ and } \sigma_i = \sqrt{\frac{1}{N-1} \sum_{k=1}^{N-1} (d^i_k - \mu_i)^2}$$

### 3.4.2 Heuristics Based on Passage Order

The estimation of specific thresholds aims at defining support sets containing the most important passages to the passage under analysis. In that sense, in this set of heuristics we explore the structure of the input source to partition the candidate passages to be in the support set in two subsets: the ones closer to the passage associated with the support set under construction, and the ones further appart.

These heuristics use a permutation, $d^i_1, d^i_2, \cdots, d^i_{N-1}$, of the distances of the passages, $s_k$, to the passage, $p_i$, related to the support set under construction, with $d^i_k = dist(s_k, p_i)$, $1 \leq k \leq N-1$, corresponding to the order of occurrence of passages $s_k$ in the input source. Algorithm 1 describes the generic procedure.

### 3.4.3 Heuristics Based on Weighted Graph Creation Techniques

There are several ways to define a weighted graph, given a dataset. The main ideia is that similar nodes must be connected by an edge with a large weight. In this set of heuristics, we explore two weight functions (Zhu, 2005) (Eqs. 24 and 25) considering that if the returned value is above a given threshold, $\delta$, the passage $s_k$ belongs to the support set of passage $p_i$, with $d^i_k = dist(s_k, p_i)$.

$$\exp(-(d^i_k - \min_{1 \leq j \leq N-1}(d^i_j))^2/\alpha^2) > \delta \qquad (24)$$

$$\left(\tanh(-\alpha(d^i_k - \frac{1}{N-1} \sum_{j=1}^{N-1} d^i_j)) + 1\right)/2 > \delta \qquad (25)$$

### 3.5 Integrating Additional Information

As argued by Wan, Yang, and Xiao (2007) and Ribeiro and de Matos (2008a), the use of additional related information helps to build a better understanding of a given subject, thus improving summarization performance. Wan et al. propose a graph-based ranking model that uses several documents about a given topic to summarize a single one of them. Ribeiro and de Matos, using the LSA framework, present a method that combines the input source

**Input**: Two values $r_1$ and $r_2$, each a representative of a subset, and the set of the passages $s_k$ and corresponding distances $d_k^i$ to the passage associated with the support set under construction
**Output**: The support set of the passage under analysis

$R_1 \leftarrow \emptyset, R_2 \leftarrow \emptyset$;
**for** $k \leftarrow 1$ *to* $N - 1$ **do**
  **if** $|r_1 - d_{i_k}| < |r_2 - d_{i_k}|$ **then**
    $r_1 \leftarrow (r_1 + d_{i_k})/2$;
    $R_1 \leftarrow R_1 \cup \{s_{i_k}\}$;
  **else**
    $r_2 \leftarrow (r_2 + d_{i_k})/2$;
    $R_2 \leftarrow R_2 \cup \{s_{i_k}\}$;
  **end**
**end**
$l \leftarrow \arg\min_{1 \leq k \leq N-1}(d_{i_k})$;
**if** $s_l \in R_1$ **then**
  **return** $R_1$;
**else**
  **return** $R_2$;
**end**

**Algorithm 1:** Generic passage order-based heuristic.

consisting of a spoken document, with related textual background information, to cope with the difficulties of speech-to-text summarization.

The model we propose may be easily expanded to integrate additional information. By using both an information source $I \triangleq p_1, p_2, ..., p_N$ and a source for additional relevant information $B$, we may redefine Eq. 12 as shown in Eq. 26 to integrate the additional information.

$$S_i \triangleq \{s \in I \cup B : sim(s, p_i) > \varepsilon_i \wedge s \neq p_i\} \tag{26}$$

Matrix $A$ (from Eq. 14) should be redefined as indicated in Eq. 27, where $a_{id_j^k}$ represents the weight of term $t_i$, $1 \leq i \leq T$ ($T$ is the number terms), in passage $p_{d_j^k}$, $1 \leq k \leq D$ ($D$ is the number of documents used as additional information) with $d_1^k \leq d_j^k \leq d_s^k$, of document $d^k$; and $a_{in_l}$, $1 \leq l \leq s$, are the elements associated with the input source to be summarized.

$$A = \begin{bmatrix} a_{1d_1^1} \dots a_{1d_s^1} & \dots & a_{1d_1^D} \dots a_{1d_s^D} & a_{1n_1} \dots a_{1n_s} \\ \dots & & & \\ a_{Td_1^1} \dots a_{Td_s^1} & \dots & a_{Td_1^D} \dots a_{Td_s^D} & a_{Tn_1} \dots a_{Tn_s} \end{bmatrix} \tag{27}$$

Given the new definition of support set and a common representation for the additional information, the most relevant content is still assessed using Eq. 13.

The same line of thought can be applied to extend the model to multi-document summarization.

## 4. Evaluation

Summary evaluation is a research subject by itself. Several evaluation models have been put forward in the last decade: beyond the long-established precision and recall (mostly useful when evaluating extractive summarization using also extractive summaries as models), literature is filled with metrics (some are automatic, others manual) like Relative utility (Radev et al., 2000; Radev & Tam, 2003), SummACCY (Hori, Hori, & Furui, 2003), ROUGE (Lin, 2004), VERT (de Oliveira, Torrens, Cidral, Schossland, & Bittencourt, 2008), or the Pyramid method (Nenkova, Passonneau, & McKeown, 2007). For a more comprehensive analysis of the evaluation field see the work by Nenkova (2006) and Nenkova et al. (2007).

Despite the number of approaches to summary evaluation, the most widely used metric is still ROUGE and is the one we use in our study. We chose ROUGE not only owing to its wide adoption, but also because one of the data sets used in our evaluation has been used in published studies, allowing us to easily compare the performance of our model with other known systems.

$$\text{ROUGE-}N = \frac{\sum_{S \in \{\text{Reference Summaries}\}} \sum_{gram_N \in S} count_{match}(gram_N)}{\sum_{S \in \{\text{Reference Summaries}\}} \sum_{gram_N \in S} count(gram_N)} \quad (28)$$

Namely, we use the ROUGE-1 score, known to correlate well with human judgment (Lin, 2004). ROUGE-$N$ is defined in Eq. 28. Moreover, we estimate confidence intervals using non-parametric bootstrap with 1000 resamplings (Mooney & Duval, 1993).

Since we are proposing a generic summarization model, we conducted experiments both in text and speech data.

### 4.1 Experiment 1: Text

In this section, we describe the experiments performed and analyze the corresponding results when using as input source written text.

#### 4.1.1 Data

The used corpus, known as TeMário, consists of 100 newspaper articles in Brazilian Portuguese (Pardo & Rino, 2003). Although our model is general and language-independent, this corpus was used in several published studies, allowing us to perform an informed comparison of our results. The articles in the corpus cover several domains, such as "world", "politics", and "foreign affairs". For each of the 100 newspaper articles, there is a reference human-produced summary. The text was tokenized and punctuation removed, maintaining sentence boundary information. Table 1 sumarizes the properties of this data set.

#### 4.1.2 Evaluation Setup

To compare the performance of our model when the input is not affected by speech-related phenomena, we use previously published state-of-the-art results for text summarization. However, since there was no information available about any kind of preprocessing for the previous studies, we could not guarantee a fair comparison of our results with the previous ones, without the definition of an adequate methodology for the comparisons.

The following systems were evaluated using the TeMário dataset:

|  |  | Average | Minimum | Maximum |
|---|---|---|---|---|
| #Words | News Story (NS) | 608 | 421 | 1315 |
|  | NS Sentence | 21 | 1 | 100 |
|  | Summary (S) | 192 | 120 | 345 |
|  | S Sentence | 21 | 1 | 87 |
| #Sentences | News Story | 29 | 12 | 68 |
|  | Summary | 9 | 5 | 18 |

Table 1: Corpus characterization.

- a set of graph-based summarizers presented by Mihalcea and Tarau (2005), namely PageRank Backward, HITS$_A$ Backward and HITS$_H$ Forward;

- SuPor-v2 (Leite, Rino, Pardo, & Nunes, 2007), a classifier-based system that uses features like the occurrence of proper nouns, lexical chaining, and an ontology;

- two modified versions of Mihalcea's PageRank Undirected, called TextRank + Thesaurus and TextRank + Stem + StopwordsRem(oval) presented by Leite et al. (2007); and,

- several complex networks summarizers proposed by Antiqueira et al. (2009).

Considering the preprocessing step we applied to the corpus and the observed differences in the published results, we found it important to evaluate the systems under the same conditions. Thus, we implemented the following centrality models:

- **Uniform Influx** (corresponds to the non-recursive, unweighted version of the model), proposed by Kurland and Lee (2005, 2010) for re-ranking in document retrieval (we experimented with several $k$ in graph definition, the sames used for support set cardinality in the $k$NN strategy, and $\mu$—10, 20, 50, 100, 200, 500, 1000, 2000, 5000, 10000—, and present only the best results);

- **PageRank**, proposed by both Mihalcea and Tarau (2004, 2005) and Erkan and Radev (2004) (passage similarity metrics differ and Mihalcea and Tarau also explore directed graphs);

- **Degree centrality** as proposed by Erkan and Radev (2004) (we experimented with several thresholds $\delta$, ranging from 0.01 to 0.09, and show only the best results); and,

- **Baseline**, in which the ranking is defined by the order of the sentences in the news article, with relevance decreasing from the begining to the end.

Table 2 further discriminates PageRank-based models. PageRank over a directed forward graph performs consistently worse (Mihalcea & Tarau, 2005) than over undirected and directed backward graphs, and it was not included in our trials. Degree and Continuous LexRank bound the performance of the LexRank model, and are the ones we use in this evaluation. Moreover, to assess the influence of the similarity metrics in these graph-based centrality models, we tested the best-performing metric of our model, the Manhattan distance, with the PageRank model. Additionally, given that the models proposed by Erkan and Radev (2004) use *idf*, we present some results (clearly identified) using both weighting schemes: using and not using *idf*.

| Proposed model | Generic designation | Similarity metric |
|---|---|---|
| Continuous LexRank | PageRank Undirected | Cosine |
| TextRank Undirected | PageRank Undirected | Content overlap |
| TextRank Backward | PageRank Backward | Content overlap |

Table 2: Models based on PageRank.

Concerning summary size, the number of words in the generated summaries directly depends on the number of words of the reference abstracts, which consisted in compressing the input sources to 25-30% of the original size.

4.1.3 Results

Table 3 illustrates the comparison between the previously proposed models and our model. In this table, our model is identified in boldface by the distance name, and the conditions used by that particular instance. Every time the best performance is achieved by an instance using supports sets whose cardinality is specified in absolute terms (1–5), we also present the best performance using support sets whose cardinality is specified in relative terms (10%–90% of the input source). For the fractional metrics, we also present the value of $N$ in Eq. 17, if $N \geq 1$, or Eq. 18, if $0 < N < 1$. For the automatically set thresholds, we identify which heuristic produced the best results using the following notation: H0 means the heuristic based on the average distance; H1 means heuristics based on the analysis of the distances progression, with H1.1 corresponding to the one based on the standard deviation, H1.2 corresponding to the one based on the diminishing differences between consecutive distances, and H1.3 corresponding to the one based on the average difference between consecutive distances; H2 means heuristics based on passage order, with H2.1 using as $r_1$ the minimum distance, and as $r_2$ the average of the distances, H2.2 using as $r_1$ the minimum distance, and as $r_2$ the maximum distance, and H2.3, using as $r_1$ the distance of the first passage and $r_2$ the distance of the second passage, according to the required permutation defined in Section 3.4.2; H3 means heuristics based on weighted graph creation techniques, with H3.1 based on Eq. 24, and H3.2 based on Eq. 25.

The best overall results were obtained by the support sets-based centrality model using both the Fractional, with $N = 1.(3)$ and using *idf*, and the Manhattan distance. The next best-performing variants of our model were Cosine, Minkowski ($N$ defined by the dimension of the semantic space), and Euclidean, all over-performing both TextRank Undirected and the Uniform Influx model. The best PageRank variant, using a backward directed graph and the cosine similarity with *idf*, achieved a performance similar to the Cosine (SSC = 4, *idf*) and the Minkowski (SSC = 2) variants of our model. TextRank Undirected, Uniform Influx, and Continuous LexRank (*idf*) obtained performances similar to the Euclidean (SSC

| Systems | ROUGE-1 | Confidence Interval |
| --- | --- | --- |
| **Fractional ($N = 1.(3)$, *idf*, H1.3)** | 0.442 | [0.430, 0.455] |
| **Manhattan (SSC = 2)** | 0.442 | [0.430, 0.454] |
| **Manhattan (10%)** | 0.440 | [0.429, 0.453] |
| **Manhattan (*idf*, H2.1)** | 0.439 | [0.428, 0.451] |
| **Cosine (*idf*, SSC = 4)** | 0.439 | [0.428, 0.451] |
| PageRank Backward Cosine (*idf*) | 0.439 | [0.427, 0.451] |
| **Minkowski (SSC = 2)** | 0.439 | [0.427, 0.452] |
| **Minkowski (H2.1)** | 0.437 | [0.426, 0.450] |
| **Cosine (*idf*, H0)** | 0.437 | [0.425, 0.449] |
| **Manhattan (H1.2)** | 0.437 | [0.425, 0.450] |
| **Euclidean (*idf*, SSC = 5)** | 0.436 | [0.424, 0.448] |
| TextRank Undirected | 0.436 | [0.424, 0.448] |
| Uniform Influx (10%NN, $\mu = 10000$) | 0.436 | [0.422, 0.449] |
| **Cosine (90%)** | 0.436 | [0.423, 0.448] |
| Continuous LexRank (*idf*) | 0.436 | [0.424, 0.448] |
| **Fractional ($N = 1.(3)$, H1.3)** | 0.435 | [0.422, 0.447] |
| **Fractional ($N = 1.(3)$, SSC = 1)** | 0.435 | [0.423, 0.448] |
| PageRank Backward Cosine | 0.435 | [0.423, 0.447] |
| Degree ($\delta = 0.02$, *idf*) | 0.435 | [0.423, 0.447] |
| TextRank Backward | 0.434 | [0.423, 0.446] |
| **Minkowski (10%)** | 0.434 | [0.422, 0.447] |
| **Euclidean (H2.3)** | 0.434 | [0.422, 0.448] |
| **Cosine (H1.3)** | 0.432 | [0.420, 0.444] |
| **Fractional ($N = 1.(3)$, 80%)** | 0.432 | [0.420, 0.445] |
| **Chebyshev (H1.2)** | 0.432 | [0.419, 0.444] |
| PageRank Backward Manhattan | 0.432 | [0.419, 0.442] |
| **Euclidean (10%)** | 0.431 | [0.418, 0.444] |
| **Chebyshev (SSC = 2)** | 0.429 | [0.417, 0.442] |
| **Chebyshev (10%)** | 0.429 | [0.417, 0.442] |
| Continuous LexRank | 0.428 | [0.415, 0.441] |
| PageRank Undirected Manhattan | 0.428 | [0.415, 0.440] |
| Baseline | 0.427 | [0.415, 0.440] |
| **Fractional ($N = 0.1$, H1.1)** | 0.427 | [0.414, 0.439] |
| Degree ($\delta = 0.06$) | 0.426 | [0.414, 0.439] |
| **Fractional ($N = 0.5$, H1.1)** | 0.422 | [0.409, 0.434] |
| **Fractional ($N = 0.75$, H1.1)** | 0.421 | [0.410, 0.433] |
| **Fractional ($N = 0.75$, 10%)** | 0.417 | [0.404, 0.429] |
| **Fractional ($N = 0.1$, 90%)** | 0.417 | [0.405, 0.429] |
| **Fractional ($N = 0.5$, 90%)** | 0.413 | [0.403, 0.425] |

Table 3: ROUGE-1 scores for the text experiment (SSC stands for Support Set Cardinality).

= 5, *idf*) and the Cosine (90%) variants. Notice that although not exhaustively analyzing the effects of term weighting, the use of *idf* clearly benefits some metrics: see, for instance, the Cosine and Fractional $N = 1.(3)$ variants of our model, the PageRank variants based on the cosine similarity, and Degree. It is relevant to note that our model, which has low computational requirements, achieves results comparable to graph-based state-of-the-art systems (Ceylan, Mihalcea, Özertem, Lloret, & Palomar, 2010; Antiqueira et al., 2009). Notice that although the estimated confidence intervals overlap, the performance of the Manhattan SCC=2 variant is significantly better, using the directional Wilcoxon signed rank test with continuity correction, than the ones of TextRank Undirected, ($W = 2584$, $p < 0.05$), Uniform Influx ($W = 2740$, $p < 0.05$), and also Continuous LexRank ($W = 2381.5$, $p < 0.1$).[4] The only variants of our model that perform below the baseline are the Fractional variants with $N < 1$. Fractional distances with $N < 1$, as can be seen by the effect of the metric on the unit circle (Figure 1), increase the distance between all passages, negatively influencing the construction of the support sets and, consequently the estimation of relevant content.

Concerning the automatically set per passage thresholds, it is possible to observe that the best overall performance was achieved by a metric, Fractional $N = 1.(3)$, with *idf*, using the heuristic based on the average difference between consecutive distances. For Cosine, Manhattan, Euclidean, and Minkowski variants, the heuristic based on the average distance (Cosine) and the heuristics based on passage order achieved results comparable to the best performing *k*NN approaches. For Chebyshev and Fractional (with $N < 1$) variants the best results were obtained using the heuristics based on the analysis of the progression of the distances.

Figure 2 shows the improvements over the baseline and over the previous best-performing system. It is possible to perceive that the greatest performance jumps are introduced by Euclidean (10%) and Euclidean (H2.3), Minkowski (SSC=2), and the best-performing Manhattan, all instances of the support sets-based relevance model. Additionally, it is important to notice that the improvement of CN-Voting over the baseline (computed in the same conditions of CN-Voting) is of only 1%, having a performance worse than the poorest TextRank version which had an improvement over the baseline of 1.6%. In what concerns the linguistic knowledge-based systems (SuPor-2 and the enriched versions of TextRank Undirected), we cannot make an informed assessment of their performance since we cannot substantiate the used baseline, taken from the work of Mihalcea and Tarau (2005). Nonetheless, using that baseline, it is clear that linguistic information improves the performance of extractive summarizers beyond what we achieved with our model: improvements over the baseline range from 9% to 17.5%. Notice however, that it would be possible to enrich our model with linguistic information, in the same manner of TextRank.

Regarding the effect of the similarity metric on the PageRank-based systems, it is possible to observe that PageRank Undirected based on Content Overlap (TextRank Undirected) has a better performance than when similarity is based on a geometric metric—either Manhattan or Cosine (Continuous LexRank). However, the same does not happen when considering the results obtained by the several variants of PageRank Backward. Although the use of Content Overlap, in fact, leads to a better performance than using a Manhattan-based similarity metric, the use of the cosine similarity results in a performance comparable to the one of using the Content Overlap metric. The Manhattan-based similarity metric is defined in Eq. 29.

---

4. Statistical tests were computed using R (R Development Core Team, 2009).

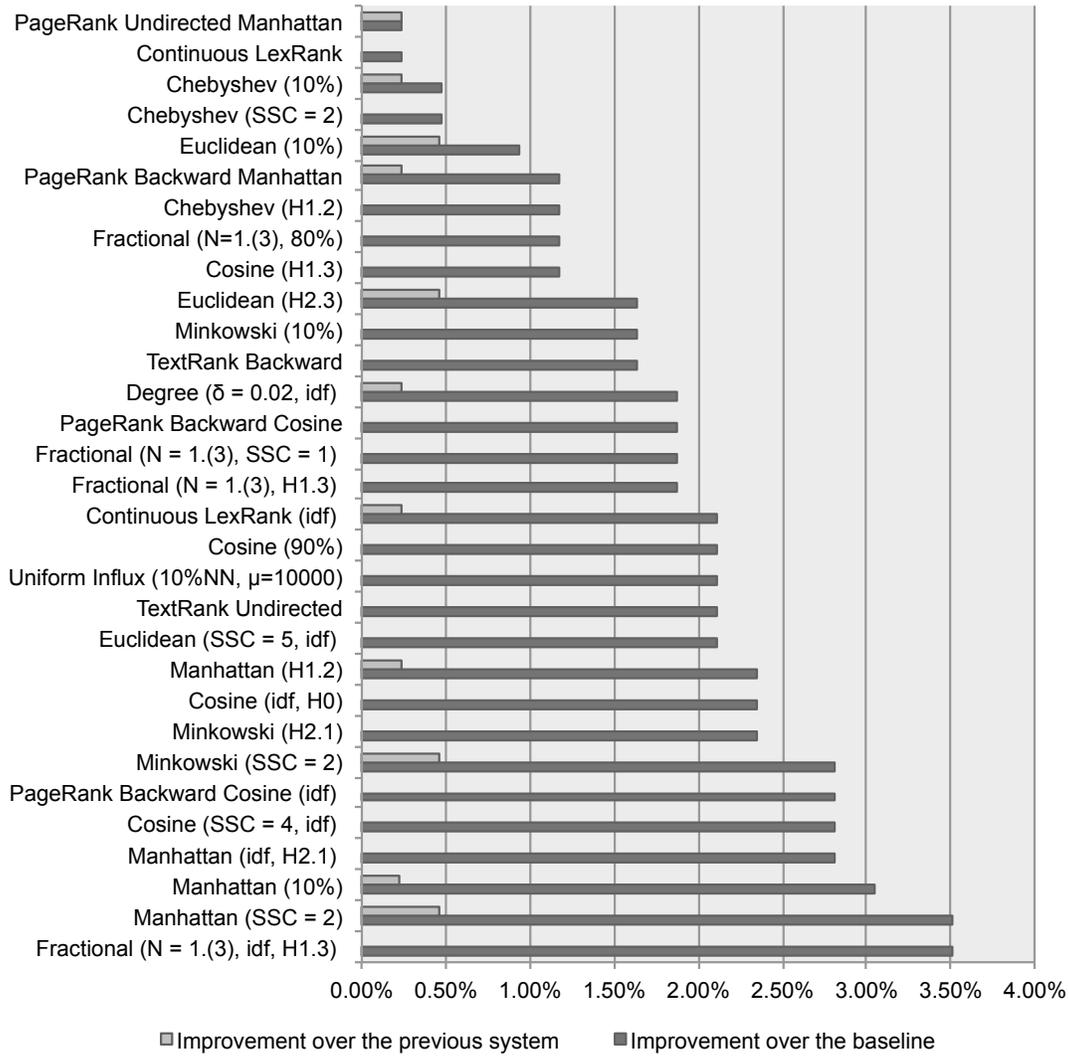

Figure 2: Analysis of the increase in performance of each model.

$$sim_{manhattan}(\mathbf{x}, \mathbf{y}) = \frac{1}{1 + dist_{manhattan}(\mathbf{x}, \mathbf{y})} \qquad (29)$$

### 4.2 Experiment 2: Speech

In this section, we describe the experiments performed and analyze the corresponding results when using as input source automatically transcribed speech.

### 4.2.1 Data

To evaluate our ideas in the speech processing setting, we used the same data of Ribeiro and de Matos (2008a): the automatic transcriptions of 15 broadcast news stories in European Portuguese, part of a news program. Subject areas include "society", "politics", "sports", among others. Table 4 details the corpus composition. For each news story, there is a human-produced reference summary, which is an abstract. The average word recognition error rate is 19.5% and automatic sentence segmentation attained a slot error rate (SER, commonly used to evaluate this kind of task) of 90.2%. As it is possible to observe in Table 4, it is important to distinguish between the notion of sentence in written text and that of sentence-like unit (SU) in speech data. Note, in particular, the difference in the average number of words per sentence in the summary versus the average number of words per SU in the news story. According to Liu, Shriberg, Stolcke, Hillard, Ostendorf, and Harper (2006), the concept of SU is different from the concept of sentence in written text, since, although semantically complete, SUs can be smaller than a sentence. This is corroborated by the fact that it is possible to find news stories with SUs of length 1 (this corpus has 8 SUs of length 1). Beyond the definition of SU, note that an SER of 90.2% is a high value: currently, the automatic punctuation module responsible for delimiting SUs achieves an SER of 62.2%, using prosodic information (Batista, Moniz, Trancoso, Meinedo, Mata, & Mamede, 2010).

|  |  | Average | Minimum | Maximum |
|---|---|---|---|---|
| #Words | News Story (NS) | 287 | 74 | 512 |
|  | NS SU | 11 | 1 | 91 |
|  | Summary (S) | 33 | 9 | 72 |
|  | S Sentence | 20 | 8 | 33 |
| #SUs | News Story | 27 | 6 | 51 |
| #Setences | Summary | 2 | 1 | 4 |

Table 4: Corpus characterization.

### 4.2.2 Evaluation Setup

Regarding speech summarization, even considering the difficulties concerning the applicability of text summarization methods to spoken documents, shallow approaches like LSA or MMR seem to achieve performances comparable to the ones using specific speech-related features (Penn & Zhu, 2008), especially in unsupervised approaches. Given the implemented models, in this experiment we compare the support sets relevance model to the following systems:

- An LSA baseline.

- The following graph-based methods: Uniform Influx (Kurland & Lee, 2005, 2010), Continuous LexRank and Degree centrality (Erkan & Radev, 2004), and TextRank (Mihalcea & Tarau, 2004, 2005).

- The method proposed by Ribeiro and de Matos (2008a), which explores the use of additional related information, less prone to speech-related errors (e.g. from online newspapers), to improve speech summarization (Mixed-Source).

- Two human summarizers (extractive) using as source the automatic speech transcriptions of the news stories (Human Extractive).

Before analyzing the results, it is important to examine human performance. One of the relevant issues that should be assessed is the level of agreement between the two human summarizers: this was accomplished using the kappa coefficient (Carletta, 1996), for which we obtained a value of 0.425, what is considered a fair to moderate/good agreement (Landis & Kosh, 1977; Fleiss, 1981). Concerning the selected sentences, Figure 3 shows that human summarizer H2 consistently selected the first $n$ sentences, and that in H1 choices there is also a noticeable preference for the first sentences of each news story.

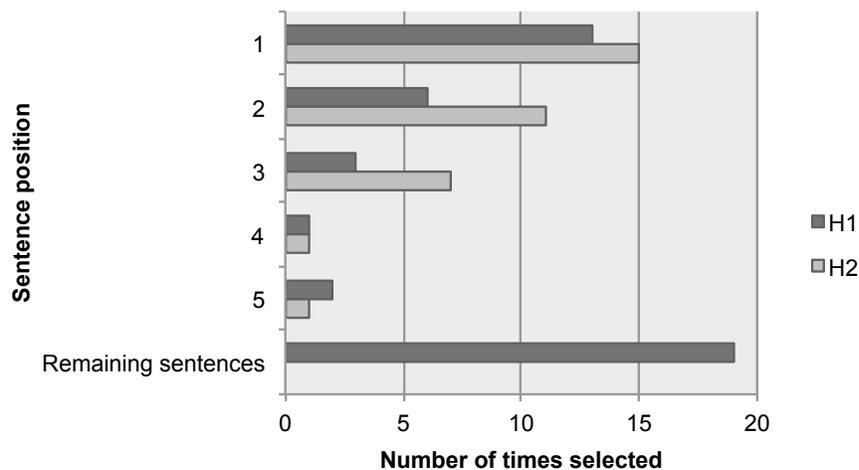

Figure 3: Human sentence selection patterns.

To be able to perform a good assessment of the automatic models, we conducted two experiments: in the first one, the number of SUs extracted to compose the automatic summaries was defined in accordance to the number of sentences of the reference human abstracts (which consisted in compressing the input source to about 10% of the original size); in the second experiment, the number of extracted SUs of the automatic summaries was determined by the size of the shortest corresponding human extractive summary. Notice that Mixed-Source and human summaries are the same in both experiments.

4.2.3 Results

Table 5 shows the ROUGE-1 scores obtained, for both speech experiments. In this table, it is possible to find more than one instance of some models, since sometimes the best-performing variant when using as summary size the size of the abstracts was different from the one using as summary size the size of the human extracts.

| Systems | Using as Summary Size Reference | |
|---|---|---|
| | Human Abstracts | Shortest Human Extracts |
| Human Extractive 1 | 0.544 [0.452, 0.640] | 0.544 [0.455, 0.650] |
| Human Extractive 2 | 0.514 [0.392, 0.637] | 0.514 [0.402, 0.652] |
| **Cosine (*idf*, H2.3)** | 0.477 [0.374, 0.580] | 0.505 [0.405, 0.619] |
| PageRank Backward Cosine | 0.473 [0.363, 0.583] | 0.510 [0.399, 0.628] |
| TextRank Backward | 0.470 [0.360, 0.580] | 0.505 [0.391, 0.625] |
| PageRank Backward Cosine (*idf*) | 0.467 [0.360, 0.571] | 0.516 [0.393, 0.646] |
| First Sentences | 0.462 [0.360, 0.572] | 0.514 [0.390, 0.637] |
| PageRank Backward Manhattan | 0.462 [0.355, 0.577] | 0.514 [0.392, 0.648] |
| **Chebyshev (H2.3)** | 0.458 [0.351, 0.571] | 0.506 [0.388, 0.618] |
| **Chebyshev (10%)** | 0.443 [0.329, 0.576] | 0.483 [0.356, 0.615] |
| **Cosine (H2.3)** | 0.410 [0.306, 0.520] | 0.446 [0.329, 0.562] |
| **Minkowski (H2.3)** | 0.407 [0.316, 0.509] | 0.449 [0.349, 0.571] |
| **Cosine (40%)** | 0.404 [0.306, 0.512] | 0.440 [0.344, 0.547] |
| **Euclidean (H2.3)** | 0.401 [0.310, 0.504] | 0.440 [0.341, 0.547] |
| Mixed-Source | 0.392 [0.340, 0.452] | 0.392 [0.339, 0.449] |
| **Cosine (*idf*, 40%)** | 0.389 [0.287, 0.500] | 0.464 [0.355, 0.577] |
| **Minkowski (40%)** | 0.381 [0.288, 0.495] | 0.435 [0.325, 0.554] |
| **Fractional ($N = 1.(3)$, *idf*, H2.3)** | 0.380 [0.274, 0.496] | 0.451 [0.354, 0.556] |
| **Manhattan (H3.1)** | 0.373 [0.276, 0.494] | 0.431 [0.343, 0.533] |
| **Fractional ($N = 1.(3)$, SSC=4)** | 0.371 [0.279, 0.483] | 0.402 [0.303, 0.526] |
| **Cosine (80%)** | 0.365 [0.290, 0.458] | 0.443 [0.345, 0.555] |
| **Fractional ($N = 1.(3)$, H3.2)** | 0.361 [0.268, 0.469] | 0.431 [0.316, 0.563] |
| Degree ($\delta = 0.06$) | 0.351 [0.247, 0.463] | 0.383 [0.276, 0.499] |
| **Fractional ($N = 1.(3)$, 20%)** | 0.347 [0.280, 0.432] | 0.374 [0.292, 0.467] |
| **Manhattan (10%)** | 0.346 [0.246, 0.478] | 0.412 [0.311, 0.532] |
| **Euclidean (20%)** | 0.344 [0.277, 0.418] | 0.371 [0.288, 0.474] |
| **Euclidean (10%)** | 0.337 [0.262, 0.432] | 0.404 [0.296, 0.524] |
| **Euclidean (SSC=3)** | 0.336 [0.262, 0.430] | 0.405 [0.300, 0.519] |
| **Fractional ($N = 1.(3)$, 10%)** | 0.336 [0.263, 0.429] | 0.405 [0.305, 0.529] |
| **Fractional ($N = 1.(3)$, SSC=3)** | 0.333 [0.256, 0.442] | 0.407 [0.308, 0.530] |
| TextRank Undirected | 0.332 [0.242, 0.423] | 0.361 [0.262, 0.464] |
| Degree ($\delta = 0.03$, *idf*) | 0.328 [0.232, 0.428] | 0.369 [0.260, 0.476] |
| Uniform Influx (10%NN, $\mu = 500$) | 0.314 [0.211, 0.427] | 0.382 [0.258, 0.511] |
| LSA Baseline | 0.308 [0.239, 0.407] | 0.338 [0.260, 0.432] |
| Continuous LexRank (*idf*) | 0.303 [0.215, 0.402] | 0.362 [0.263, 0.471] |
| **Fractional ($N = 0.1$, 90%)** | 0.301 [0.191, 0.438] | 0.368 [0.252, 0.498] |
| Continuous LexRank | 0.279 [0.212, 0.343] | 0.335 [0.246, 0.441] |
| PageRank Undirected Manhattan | 0.234 [0.163, 0.295] | 0.328 [0.240, 0.432] |
| **Fractional ($N = 0.5$, 90%)** | 0.224 [0.133, 0.336] | 0.284 [0.176, 0.412] |
| **Fractional ($N = 0.75$, 90%)** | 0.208 [0.149, 0.281] | 0.235 [0.165, 0.302] |

Table 5: ROUGE-1 scores, with 95% confidence intervals computed using bootstrap statistics, for the speech experiment (SSC stands for Support Set Cardinality; sorted using the scores of the human abstracts).

295

A first observation concerns a particular aspect of corpus: as it can be seen, especially in the experiment using as reference size the size of the shortest human extracts, both Human 2 and First Sentences summarizers attained the same ROUGE-1 scores (this does not happen in the experiment using the abstracts size only, due to the fact that First Sentences summaries are shorter, adapted to the experiment required size, than the ones of Human 2, which were not changed). In fact, the summaries are equal, which shows a consistent bias indicating that the most relevant sentences tend to occur in the beginning of the news stories. This bias, although not surprising, since the corpus is composed of broadcast news stories, is also not that common as can be seen in previous work (Ribeiro & de Matos, 2007; Lin, Yeh, & Chen, 2010). Second, it is interesting to notice the performance of the PageRank-based models: while in text there is no observable trend concerning the directionality of the graph, and both LexRank versions performed above the baseline, in speech only the backward versions achieved a good performance (the four undirected versions performed around the baseline, with LexRank obtaining results below the LSA baseline, with exception for the experiment using the extracts size and *idf*). From a models perspective, and considering the performance of backward versions in both text and speech, the use of backward directionality seems the main reason for the good performance in speech, where input sources consist of transcriptions of broadcast news stories from a news program. In fact, as mentioned before, this kind of input source is usually short (cf. Table 4) and the main information is given in the opening of the news story. This suggests that directionality introduces position information in the model, which is only relevant for specific types of input source (this is also discussed in Mihalcea & Tarau, 2005). Moreover, note that Continuous LexRank performance was close to the LSA Baseline, which implies that the model is quite susceptible to the referred bias, to the noisy input, or to both. Taking into consideration that the model is based on pair-wise passage similarity and that one of the best-performing support sets-based instance was Cosine, the same similarity metric used by LexRank, it seems that the model was not able to account for the structure of the input sources of this data set. In fact, Degree centrality, also based on the cosine similarity performed better than all PageRank Undirected models. The Influx model performed close to Degree centrality, far from the best performing approaches, which, in this case, suggests that the method for generating the graph, the generation probabilities, is affected by the noisy input, especially when considering small contexts like passages. Approaches based on generation probabilities seem more adequate to larger contexts, such as documents (Kurland & Lee, 2005, 2010; Erkan, 2006a). Erkan (2006b) mentions that results in query-based summarization using generation probabilities were worse than the ones obtained by LexRank in generic summarization.

Concerning the overall results, performance varies according to the size of the summaries. When using the abstracts size, the best-performing instance is Cosine with *idf* using an heuristic based on the passage order; when using the reference extracts size, the best performance was achieved by the backward PageRank model, followed by the Chebyshev variant also using an heuristic based on passage order and the same Cosine variant. Both variants achieved better results than TextRank Backward. Given the success of the heuristic H2.3 in these experiments, it seems that this heuristic may also be introducing position information in the model. Although not achieving the best performance in the experiment using the extracts size, there is no significant difference between the best sup-port sets-based relevance model instance, the Chebyshev variant using an heuristic based on passage order, and the ones achieved by human summarizers: applying the directional Wilcoxon signed rank test with continuity correction, the test values when using the shortest human extracts size are $W = 53$, $p = 0.5$. This means a state-of-the-art performance in the experiment using the abstracts size, and comparable to a human (results similar to First Sentence, which is similar to Human Extractive 2) when using the shortest human extracts size. In fact, Chebyshev (10%), to avoid the influence of possible position information, is also not significantly different than Human Extractive 2 ($W = 11$, $p = 0.2092$). Cosine with *idf* and using H2.3 has a better performance with statistical significance than Degree with $\delta = 0.06$ ($W = 53.5$, $p < 0.005$ when using the abstracts size; $W = 54$, $p < 0.005$ when using the shortest human extracts size), TextRank Undirected ($W = 92.5$, $p < 0.05$ when using the abstracts size; $W = 96$, $p < 0.05$ when using the shortest human extracts size), and Uniform Influx ($W = 60$, $p < 0.01$ when using the abstracts size; $W = 51$, $p < 0.06$ when using the shortest human extracts size), using the same statistical test. The obtained results, in both speech transcriptions and written text, suggest that the model is robust, being able to detect the most relevant content without specific information of where it should be found and performing well in the presence of noisy input. Moreover, cosine similarity seems to be a good metric to use in the proposed model, performing among the top ranking variants, in both written and spoken language.

Fractional variants with $N < 1$ were, again, the worst performing approaches (we did not include values for the automatically set per passage thresholds in Table 5, since they were worse than the simple *k*NN approach) because their effect on the similarity assessment boosts the influence of the recognition errors. On the other hand, Chebyshev seems more imune to that influence: the single use of the maximal difference through all the dimensions makes it less prone to noise (recognition errors). The same happens with the variant using the generic Minkowski distance with $N$ equal to the number of dimensions of the semantic space.

Figures 4 and 5 shows the performance variation introduced by the different approaches. Notice that, in the speech experiments, performance increments are a magnitude higher when compared to the ones in written text. Overall, the Chebyshev variant of the support sets-based relevance model introduces the highest relative gains, close to 10% in the experiment using the abstracts size, close to 5% in the experiment using the extracts size. In the experiment using the extracts size, TextRank Undirected also achieves relative gains of near 10% over the previous best-performing system, the LSA baseline. Similar relative improvements are introduced by the human summarizers in the experiment using the abstracts size. As expected, increasing the size of the summaries increases the coverage of the human abstracts (bottom of Figure 5).

Further, comparing our model to more complex (not centrality-based), state-of-the-art models like the one presented by Lin et al. (2010) suggests that at least similar performance is attained: the relative performance increment of our model over LexRank is of 57.4% and 39.8% (both speech experiments), whereas the relative gain of the best variant of the model proposed by Lin et al. over LexRank is of 39.6%. Note that this can only be taken as indicative, since an accurate comparison is not possible because data sets differ, Lin et al. do not explicit which variant of LexRank is used, and do not address statistical significance.

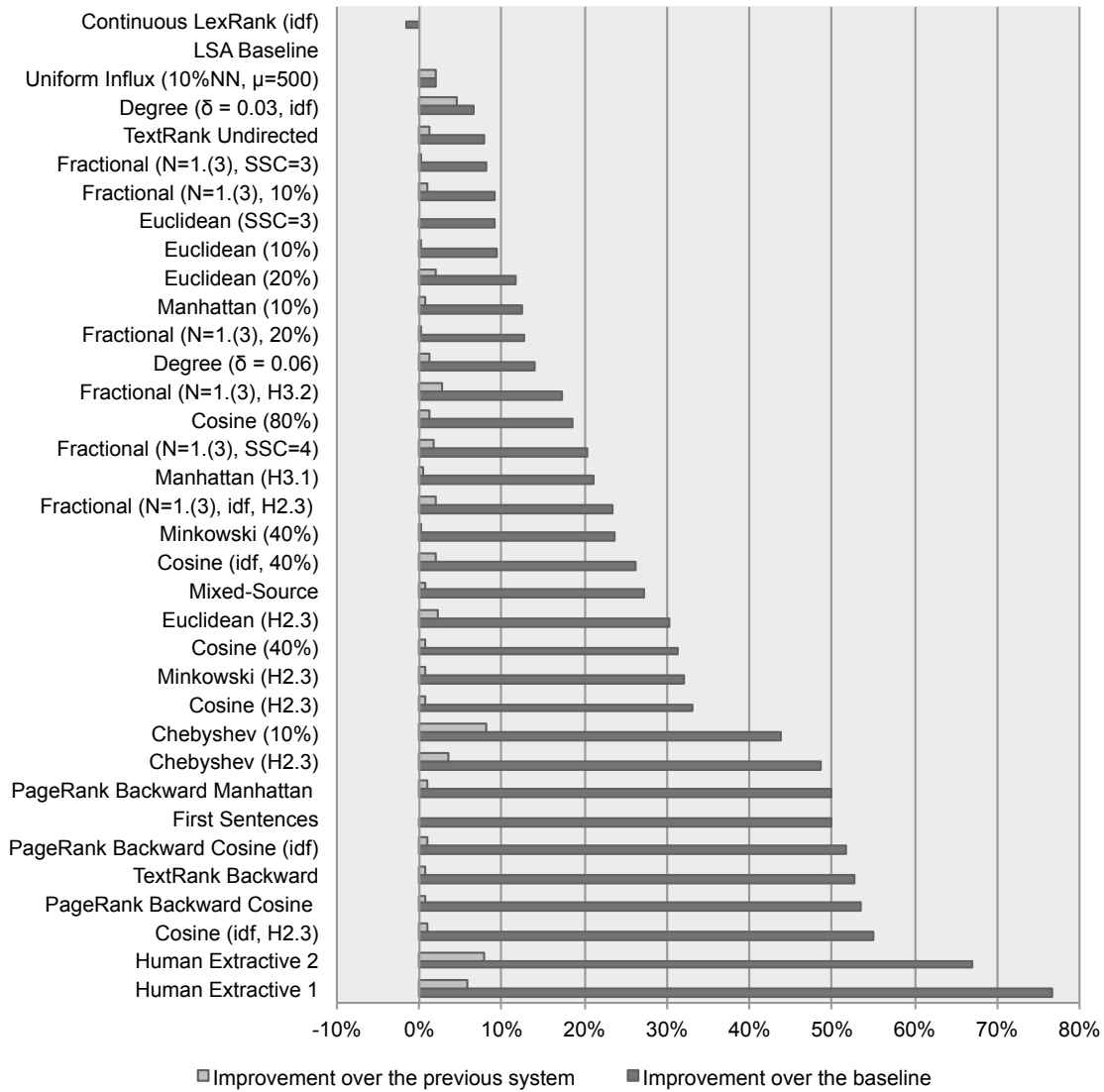

Figure 4: Analysis of the increase in performance of each model (Experiment using the abstracts size).

### 4.3 Influence of the Size of the Support Sets on the Assessment of Relevance

We do not propose a method for determining an optimum size for the support sets. Nonetheless, we analyze the influence of the support set size on the assessment of the relevant content, both in text and speech.

Figure 6 depicts the behavior of the model variants with a performance above the baseline over written text, while Figure 7 illustrates the variants under the same conditions over automatic speech transcriptions (in this case, error bars were omitted for clarity). We analyze the general performance of those variants, considering as support set size the number of passages of the input source in 10% increments. Given the average size of an input source, both in written text (Table 1), and speech transcriptions (Table 4), absolute cardinalities (SSC) ranging from 1 to 5 passages broadly cover possible sizes in the interval 0-10%.

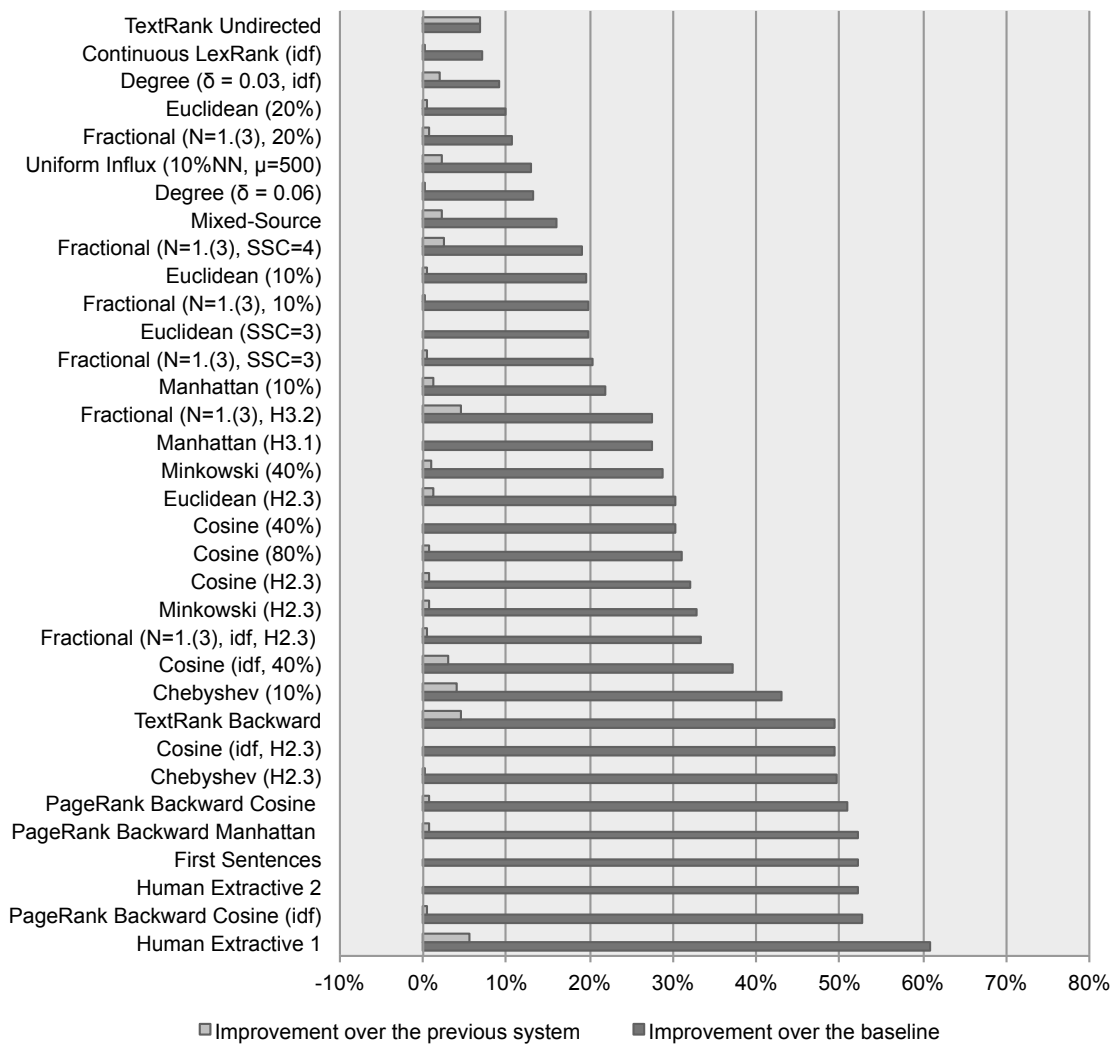

Figure 5: Analysis of the increase in performance of each model (Experiment using the abstracts size).

A first observation concerns the fact that varying the cardinality of the support sets when the input sources consist of written text has a smooth effect over the performance. This allows the analysis of generic tendencies. In contrast, when processing automatic speech transcriptions, it is possible to perceive several irregularities. These irregularities can have two different causes: the intrinsic characteristics of speech transcriptions such as recognition errors, sentence boundary detection errors, and the type of discourse; or, the specificities of the data set—in particular, the global size of the corpus and the specific news story structure. However, considering the performance of the metrics over both text and speech, the irregularities seem to be mainly caused by the intrinsic properties of speech transcriptions and the specific structure of the news story.

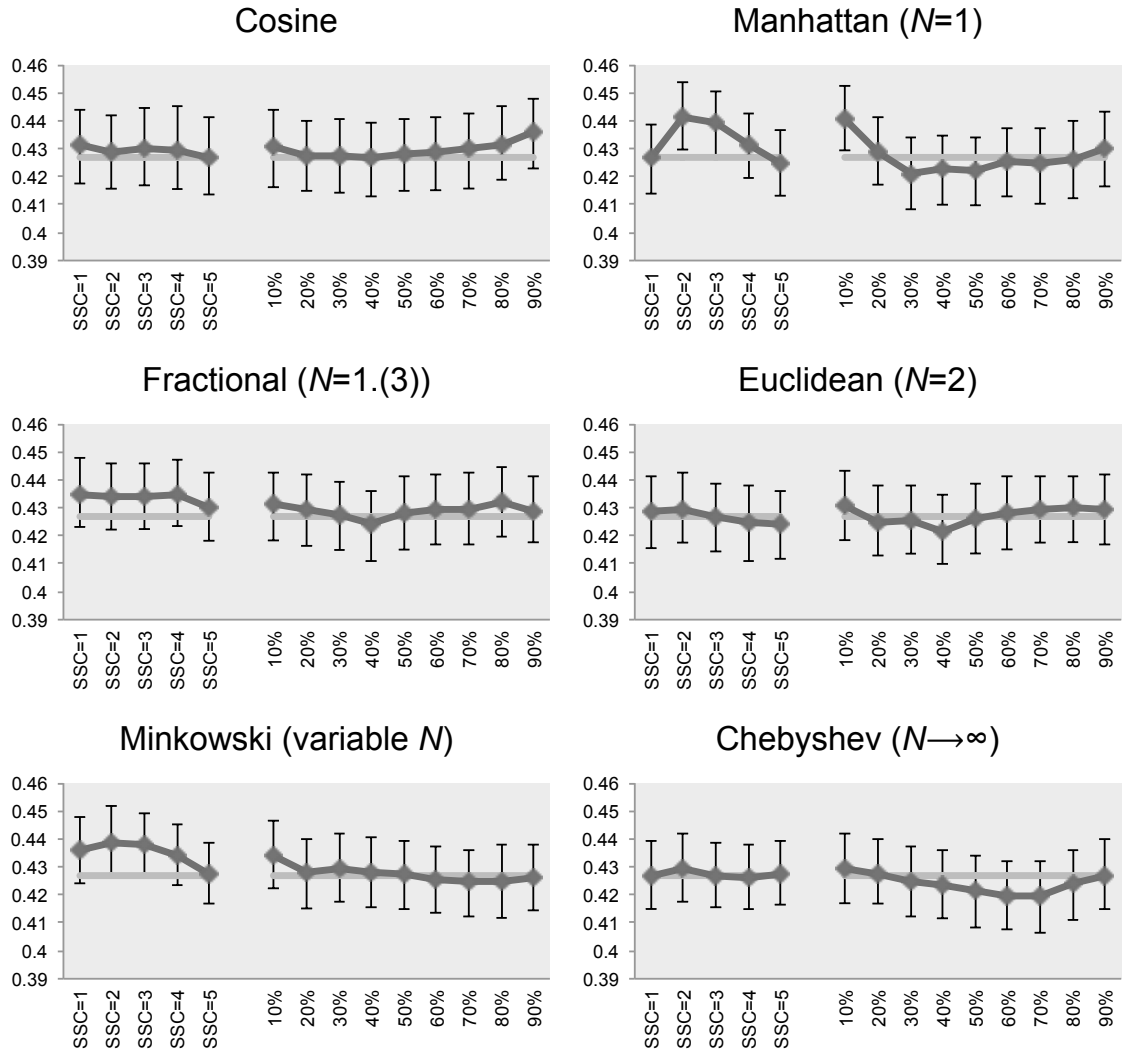

Figure 6: Analysis of the impact of the cardinality of the support sets over text summarization. Y axes are ROUGE-1 scores and X axes are support sets cardinalities (absolute and relative to the length of the input source, in terms of passages).

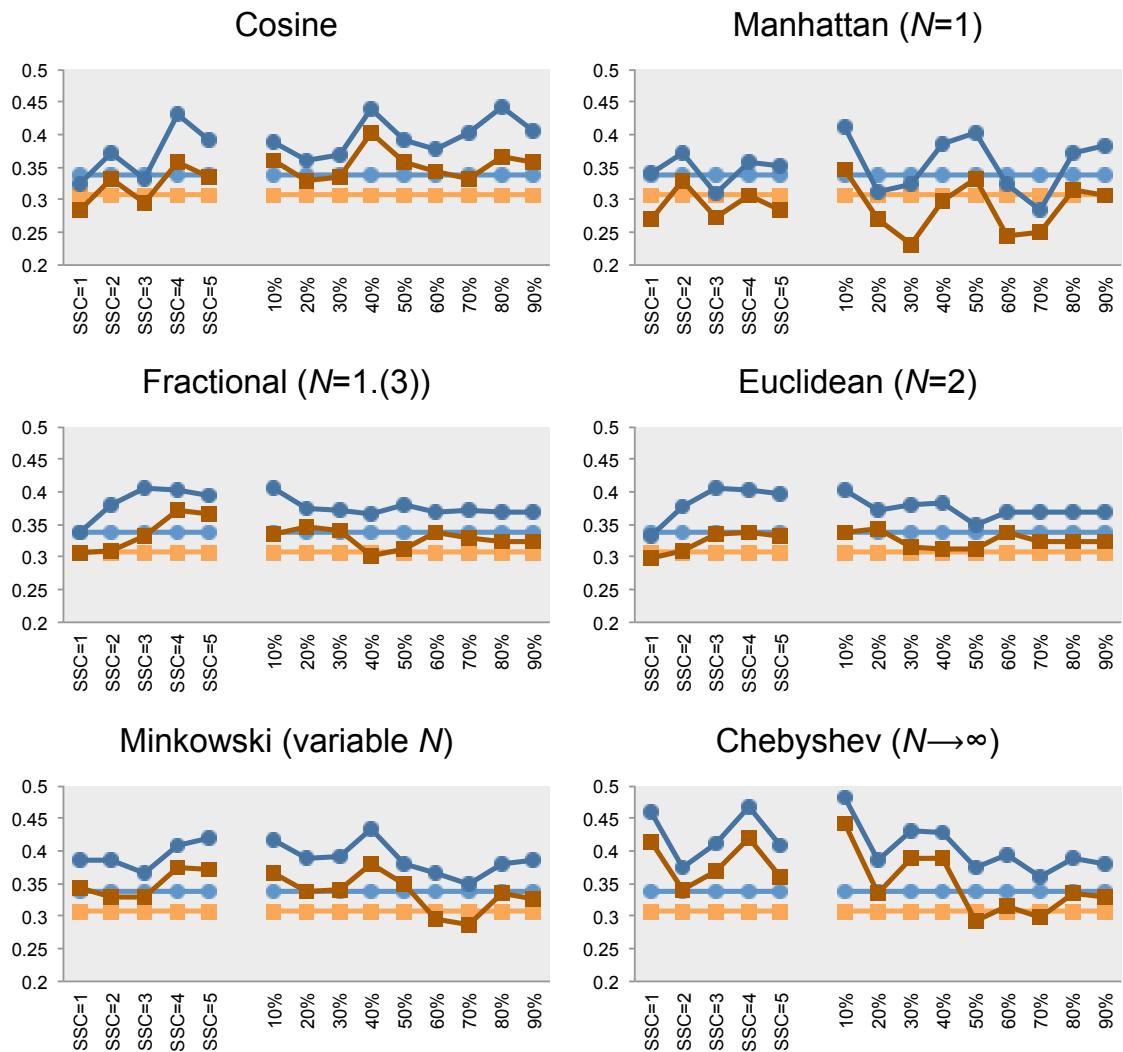

Figure 7: Analysis of the impact of the cardinality of the support sets over speech-to-text summarization. Y axes are ROUGE-1 scores and X axes are support sets cardinalities (absolute and relative to the length of the input source in terms of passages). Lines with square marks correspond to the experiment using as summary size the size of human abstracts; lines with circle marks correspond to the experiment using as summary size the size of the shortest human extracts. Horizontal lines correspond to baselines.

Concerning performance itself, in written text, the best performances are achieved using low cardinalities (absolute cardinalities of 2 or 3 passages, or of about 10% of the passages of the input source). Moreover, an increase in the size of the support sets leads to a decay of the results (except when using the cosine similarity). When processing automatic speech transcriptions, it is difficult to find a clear definition of when the best results are achieved. Considering absolute cardinalities, with the exception of the Manhattan distance, every variant has a peak when using support sets with 4 passages. However, it is not possible to extend such line of thought to relative sizes due to the previously referred irregularities. Nonetheless, higher cardinalities (70%–90%) lead to worse results, what is expected given the nature of the model (again with exception of when using the cosine similarity). In addition, note that increasing the size of the summaries improves the distinction from the baseline (summaries based on the size of the shortest human extracts are longer than the ones based on the size of the human abstracts). This means that the model is robust to needs regarding summary size, continuing to select good content even for larger summaries.

## 5. Conclusions

The number of up-to-date examples of work on automatic summarization using centrality-based relevance models is significant (Garg, Favre, Reidhammer, & Hakkani-Tür, 2009; Antiqueira et al., 2009; Ceylan et al., 2010; Wan, Li, & Xiao, 2010). In our work, we assessed the main approaches of the centrality-as-relevance paradigm, and introduced a new centrality-based relevance model for automatic summarization. Our model uses support sets to better characterize the information sources to be summarized, leading to a better estimation of the relevant content. In fact, we assume that input sources comprehend several topics that are uncovered by associating to each passage a support set composed by the most semantically related passages. Building on the ideas of Ruge (1992), *[...] the model of semantic space in which the relative position of two terms determines the semantic similarity better fits the imagination of human intuition [about] semantic similarity [...]*, semantic relatedness was computed by geometric proximity. We explore several metrics and analyze their impact on the proposed model as well as (to a certain extent) on the related work. Centrality (relevance) is determined by taking into account the whole input source, and not only local information, using the support sets-based representation. Moreover, although not formally analyzed, notice that the proposed model has low computational requirements.

We conducted a thorough automatic evaluation, experimenting our model both on written text and transcribed speech summarization. The obtained results suggest that the model is robust, being able to detect the most relevant content without specific information of where it should be found and performing well in the presence of noisy input, such as automatic speech transcriptions. However, it must be taken into consideration that the use of ROUGE in summary evaluation, although generalized, allowing to easily compare results and replicate experiments, is not an ideal scenario, and consequently, results should be corroborated by a perceptual evaluation. The outcome of the performed trials show that the proposed model achieves state-of-the-art performance in both text and speech summarization, including when compared to considerably more complex approaches. Nonetheless, we identified some limitations. First, although grounding semantic similarity on geometric proximity, in the current experiments we rely mainly on lexical overlap. While maintaining the semantic approach, the use of more complex methods (Turney & Pantel, 2010) may improve the assessment of semantic similarity. Second, we did not address a specific procedure for estimating optimum thresholds, leaving it for future research. Nonetheless, we explored several heuristics that achieved top ranking performance. Moreover, we carried out in this document an analysis that provides some clues for the adequate dimension of the support sets, but a more analytical analysis should be performed.

## Acknowledgments

We would like to thank the anonymous reviewers for their insightful comments. This work was supported by FCT (INESC-ID multiannual funding) through the PIDDAC Program funds.